\documentstyle[12pt]{article}
\def\Zop{{\bf Z}}

\newcommand{\CP}{Chan Paton\ }

\newcommand{\II}{{{\cal I}_4}}
\newcommand{\D}{{\cal D}}
\newcommand{\C}{{\cal C}}

\newcommand{\PP}{{\cal P}}

\newcommand{\wt}{\widetilde}
\newcommand{\wh}{\widehat}

\newcommand{\ba}{{\bf a}}
\newcommand{\bb}{{\bf b}}
\newcommand{\bc}{{\bf c}}
\newcommand{\bk}{{\bf k}}
\newcommand{\bm}{{\bf m}}
\newcommand{\Z}{{\bf Z}}
\newcommand{\tq}{{q}}
\newcommand{\ta}{{\theta}}

\newcommand{\be}{\begin{equation}}
\newcommand{\ee}{\end{equation}}
\newcommand{\ben}{\begin{eqnarray}\displaystyle}
\newcommand{\een}{\end{eqnarray}}

\newcommand{\sectiono}[1]{\section{#1}\setcounter{equation}{0}}

\begin{document}
\begin{titlepage}
\rightline{BROWN-HET-1205}
\rightline{hep-th/9910}
\vskip 1cm
\centerline{ \Large\bf{Non-BPS Branes on a Calabi-Yau Threefold and }}
\vskip 1cm
\centerline{\Large\bf{Bose-Fermi Degeneracy}}
\vskip 1cm
\centerline{\sc Mihail Mihailescu$^{b}$$^{1}$, Kyungho Oh$^{a}$ and Radu
Tatar$^{b}$$^{2}$}
\vskip 1cm
\centerline{$^a$ Dept. of Mathematics, University of 
California, Santa Barbara, CA 93106, USA }
\centerline{and}
\centerline{ Dept. of Mathematics, University of 
Missouri-St. Louis, St. Louis, MO 63121, USA}
\centerline{{\tt oh@math.umsl.edu}}
\centerline{$^b$ Dept. of Physics, Brown University,
Providence, RI 02912, USA}
\centerline{{\tt $^1$ mm@het.brown.edu}}
\centerline{{\tt $^2$ tatar@het.brown.edu}}
\vskip 2cm
\centerline{\sc Abstract}
We consider the spectrum of open strings for non-BPS D-brane configuration
in type II string theory on a Calabi-Yau threefold. In general, there is
no degeneracy between bosonic and  fermionic states. However we find special 
values for the
moduli space of Calabi-Yau threefolds there are
 non-BPS brane configurations which have an exact degeneracy between
bosonic and fermionic states. For these values there is no force between
pairs of non-BPS D-branes. This gives rise to a possibility of building
diverse non-supersymmetric gauge field theories on the brane world-volume.
We use  the approach
recently elaborated by Gaberdiel and Sen.

 \def\today{\ifcase\month\or
January\or February\or March\or April\or May\or June\or
July\or August\or September\or October\or November\or December\fi,
\number\year}
\vskip 1cm
\end{titlepage}
\newpage
\sectiono{Introduction}
Our current understanding of string theory suggests that there are five consistent
superstring models: type IIA, type IIB, type I , SO(32) and 
$E_8 \times E_8$ heterotic and a unique eleven dimensional 
supergravity theory, all
of them being perturbative expansions of an underlying theory , M
theory. We know how to connect various corners of M theory moduli space
corresponding to different string theories but we do not have a clear picture
of what happens in the intermediate region. 

The main tool to connect various theories are  D-branes which are
solitons with the property that the open strings can end on them. 
They are supersymmetric solutions of the equations of motion and sit in
the short representations of the supersymmetry algebra (BPS states).  The
last property
allows the D-branes to   remain stable when one goes from a weak to a strong
coupling constant. 

An important question is what happens when we try to analyze the
non-perturbative duality beyond the BPS level. Are the non-BPS states
stable? If yes, how do we map them in duality?

The existence  of stable  non-BPS D-branes has significant ramifications on the gauge field
theories on the D-brane world-volume.  As in the case of
supersymmetric field theories studied
from supersymmetric brane configurations (for a detailed review see
\cite{gk}),  non-supersymmetric gauge field theory can be studied from
the  non-BPS D-branes if there is no interacting  force between them.

The study of non-BPS states has started with the work of Sen
\cite{se1,se2,se3,se4,se5,sen1} who built non-BPS 
D-branes by considering pairs of branes-antibranes and applying an
orbifold GSO projection. In \cite{se4,se5} the heterotic/Type I
duality was tested at  non-BPS level by identifying non-BPS and stable
states which are mapped into each other.  
In \cite{og1,og2,og3,og4,ga,dst}, the D-branes were considered as boundary states
in
the closed string theory that satisfy different projections and
conservation conditions.
For detailed reviews of non-BPS D-branes and their use see 
\cite{lr,sen,sch}. When the BPS D-branes are regarded as tachyonic kinks 
 solutions on non-BPS D-branes of higher dimensions, descent
relations between BPS and non-BPS states are obtained and this allows to
identify the
D-brane charges with elements of K-theory as in
\cite{k1,k2,k3,k4,k5,k6,k7,k8}.
 
In \cite{sen2},  
non-supersymmetric systems of D-branes in type IIA/IIB string
theory compactified on an orbifold K3 were considered. For special points
in the moduli
space of the K3 orbifold an exact Bose-Fermi degeneracy in
the open string spectrum was obtained. This means that up to one loop in
open string theory, the D-branes do not exert any force between them. 

In this paper we go one step further by considering non-supersymmetric
systems of D-branes  compactified
on a Calabi-Yau threefold  
in  type IIA/IIB string theory. This will give rise to a four dimensional
non-supersymmetric theory.
The Calabi-Yau threefold considered in this paper is a quotient of a
 product of the 
K3 orbifold and a two dimensional torus. This particular Calabi-Yau
orbifold has been considered in  
\cite{fhsv, sen1}.
The D-branes we consider wrap cycles of this Calabi-Yau threefold
and are also extended in some of the  non-compact directions.
In T-dual description, these D-branes wrap 
 non-supersymmetric cycles on a  Calabi-Yau threefold.
By using these brane configuration, we are going to study the one-loop
partition function with open strings and the tree-level partition function
with closed strings. We  show that both these partition functions
vanish
at  special points of the
moduli space of Calabi-Yau threefold.  This implies  that the branes do not
exert any force between them. 

The content of this paper is as follows: in section 2 we describe
 a  Calabi-Yau threefold that we are interested in and its corresponding
cycles. In Section 3 we describe the non-BPS states and we calculate
the one-loop partition function by using open strings ending on the
corresponding non-BPS D-branes. In section 3 we calculate 
the tree level partition function from
a closed string approach. We find
 the critical radii of the compact directions for which 
the partition function vanishes.

\sectiono{Cycles on a Calabi-Yau threefold} \label{s1}
\setcounter{equation}{0}
We consider D branes wrapped on non-supersymmetric 2-cycle or 3-cycle of a
Calabi-Yau threefold. The Calabi-Yau threefold we are concerned is
a quotient of  $T^{6}$ by  $\Z_{2} \times \Z_{2}$.
This was considered  in \cite{fhsv, sen1}.
Let  $x^4, \ldots , x^9$ be the coordinates of $T^6$ with radii
$R_4, \ldots , R_9$.
The action of $\Z_{2} \times \Z_{2}$ will be generated by the
actions $\II, \II'$ such that 
\ben
\label{first}
\II: (x^4,\ldots x^9)&\to& (x^4, x^5, -x^6, -x^7, -x^8, -x^9)\\
\label{second}
\II': (x^4,\ldots x^9)&\to& (-x^4,-x^5,-x^6+\pi R_6,-x^7,x^8+\pi
R_8,x^9)\, .
\een
Thus we compactify type IIA string theory $T^6$ and mod out the
theory by the $\Z_{2} \times \Z_{2}$ symmetry.
By modding out by  the first action $\II$ on $T^6$, we obtain
$T^2 \times T^4/\Z_2$. There are 16 fixed points on $T^4$ under $\Z_2$
induced by $\II$. Thus there are 16 tori  which form
the singular locus of $T^2 \times T^4/\Z_2$. By blowing up 
the 16 fixed points on $T^4/\Z_2$, 
we obtain a product of a torus $T^2$ and a K3 surface. 
Now we return to the
singular space $T^2 \times T^4/\Z_2$ and note that
$\II'$ induces an involution on $T^2 \times T^4/\Z_2$ without fixed points.
This is
because
$x^8$-coordinates are shifted by $\pi R_8$. By taking a further quotient of
$T^2 \times T^4/\Z_2$ by $\II'$, we obtain a Calabi-Yau orbifold.
The singular locus on
 this Calabi-Yau orbifold is the images of 16 fixed tori under
$\II'$. Thus the singular locus consists of 8 tori.

To construct a non-BPS D-brane configuration on this Calabi-Yau orbifold,
we have to begin with a non-BPS D-brane configuration wrapping on
cycles on $T^6$ which are
invariant under $\II$ and $\II'$. The $\II$ and $\II'$ invariant
cycles on $T^6$ are the images of $\II'$ invariant cycles on  
$T^2 \times T^4/\Z_2$
under the quotient map $\II'$.
We may obtain $\II'$ invariant two or three 
cycles on $T^2 \times T^4/\Z_2$ in the 
following manner. Let $C$ be  a 1-cycle on $T^2$ and 
$S$ be  a 2-cycle on $T^4/\Z_2$.
We denote the images of $C$ under $\II'$  by $C'$ and 
the images of $S$ under $\II'$ by $S'$ respectively.
Then
$S + S'$ and $C \times S + C' \times  S'$
are $\II'$ invariant two and 
three cycles on $T^2 \times T^4/\Z_2$ respectively.
Then by taking a quotient by $\II'$ we obtain
cycles on the Calabi-Yau orbifold $T^2 \times T^4/\Z_2 \times \Z_2$.

\sectiono{Open String Approach}
We want to study the case of a non-BPS state of type IIA string theory on
the Calabi-Yau orbifold constructed in $\S 2$. 
In order 
to have a non-BPS D-brane in type IIA on the Calabi-Yau orbifold,
we need to have an odd number of
tangential directions of the D-brane along $T^6$.\footnote{We would
like to thank professor Ashoke Sen for explaining to us this and other
details of this section.} Then we take a union of  a non-BPS D-brane 
and its transformation under $\II'$ action.

More specifically,  we start with  a D1
string wrapped along the  compact $x^9$-th direction  located at the fixed points
of the action $\II$ which, after
T-duality, can be  identified with a D-brane wrapped on  a non-supersymmetric
2-cycle. In general, 
the p-branes we deal with in this paper are  extended
along the non-compact directions except the $x^9$-th direction.
The non-compact directions are  transverse to the Calabi-Yau orbifold.
Thus  we have a $p$-brane $D$ at
\be \label{ecy1}
x^4=x^5=x^6=x^7=x^8=0\, ,
\ee
and its transformation $D'$ at
\be \label{ecy2}
x^4=x^5=0, x^6=\pi R_6, \quad x^7=0, \quad x^8=\pi R_8\, .
\ee
The pair $D, D'$ gives a non-BPS D-brane configuration on the $T^6$ 
which is invariant under
$\II$ and $\II'$. 
 We will also discuss the case 
$x_4 \ne 0, x_5 \ne 0$
in the second part of this section. 

Now We will  calculate the open string partition function and show it
vanishes at the critical radii in $(x^6, x^7, x^8, x^9)$ direction:
\be \label{openpart1}
Z = \int {dt\over 2t}\, Tr_{NS-R} (e^{- 2 t H_o} \PP)\, ,
\ee
where NS and R denote Neveu-Schwarz and Ramond sectors respectively.
$\PP$ is a projection operator which ensures that the Fock space is
physical and $H_o$ is the open string Hamiltonian:
\be \label{openhamilt}
H_o= \pi {\vec p}^2 + {1\over 4\pi} {\vec w}^2
+ \pi \sum_{\mu =0,3,\ldots 9}  
[\sum_{n=1}^\infty \alpha^\mu_{-n} \alpha^\mu_n + \sum_{r>0} r
\psi^\mu_{-r} \psi^\mu_r] + \pi C_o\, ,
\ee
where $\vec p$ denotes the open string momentum along the 
directions for which the string has Neumann (N) boundary
conditions at both ends, and $\vec w$ denotes the winding
modes along the directions for which both ends obey Dirichlet (D)
boundary conditions. $\alpha^\mu_n$ and $\psi^\mu_r$ are the 
bosonic and fermionic oscillators satisfying the usual commutation and
anti-commutation relations.
The index $n$ always will be integer values, 
whereas the index $r$ will be integer (integer + 
${1\over 2}$) values in the R (NS) sector
for directions satisfying the same boundary condition at both ends of
the open string ({\it i.e.} both Neumann (N) or both Dirichlet (D)). 
For
directions satisfying different boundary conditions at the two ends
of the open string (one D and one N) the index $n$ will be
 integer+${1\over 2}$
values and the index $r$ will be 
 integer +${1\over 2}$ (integer) values in the R
(NS) sector. The normal ordering constant $C_o$ vanishes in the
R-sector and is equal to $-{1\over 2}+{s\over 8}$ in the NS sector (in
$\alpha'=1$ units) where $s$ denotes the number of coordinates
satisfying D-N boundary conditions.

For each D brane we
have two different Chan-Paton factors - the $2 \times 2$ identity matrix I
and the Pauli matrix $\sigma_{1}$. Because we have 2 D-branes, there are 
four \CP sectors and  these four
sectors are $D D, D' D', D D', D' D$ respectively. 
In order to calculate the open string partition function, we need to
calculate the contribution from sectors which are invariant under
$\II'$ 
But the action of 
$\II'$  just  exchanges the $D D$ sector with $D' D'$ sector and the 
$D D'$ sector with the $D' D$ sector so that its action  just  reduces the
number of open strings by half.
Thus it is enough to consider the contribution from the $D D'$ and $D D$
sectors.
  As discussed in 
\cite{sen2} the combined contribution of the Chan Paton factors from NS
sector states is:
\be
\label{c1}
\int {dt\over 2t} tr_{NS} \left(e^{- 2 t H_o} {1 + (-1)^F\cdot g\over 2} 
\right)
\ee
and the contribution
from the R sector is 
\be
\int {dt\over 2t} {1\over 2} \, tr_{R} (e^{- 2 t H_o})
\label{c2}
\ee
where all the traces are over the Fock space of oscillators.
We introduce some functions to simplify the presentation,
\ben
\tq& =&e^{-\pi t}\, ,\\
\ta (R)& = &\sum_{n\in  \Zop} \tq^{2R^2 n^2}\, .
\een
We also introduce Dedekind $\eta$-type functions\, ,
\ben
f_1 (\tq ) &=& \tq^{{1\over 12}} \prod_{n=1}^{\infty} (1- \tq^{2n})\, ,\\
f_2 (\tq ) &= & \sqrt{2} \tq^{{1\over 12}} \prod_{n=1}^{\infty} (1+ \tq^{2n})\, ,\\
f_3 (\tq)  &= & \tq^{-{1\over 24}} \prod_{n=1}^{\infty} (1+ \tq^{2n-1})\, ,\\
f_4 (\tq) &= & \tq^{-{1\over 24}} \prod_{n=1}^{\infty} (1- \tq^{2n-1}).
\een
Before calculating the contributions to the partition function we need to
make some observations. The $D$ brane is located at
 $x^6 = 0, x^8 = 0$, the
$D'$ brane is located  at $x^6 = \pi R_6, x^8 = \pi R_8$, and the distance
between them is  $\pi\sqrt{(R^6)^2
+ (R^8)^2}$. To take care of  the winding
modes computation of the open string properly, one needs to choose
new rotated  coordinates ${x^6}', {x^8}'$ for the $(x^6, x^8)$ plane.
The new coordinates will be given by
\ben
{x^6}' = R_8 x^6 - R_6 x^8,\quad
{x^8}' = R_6 x^6 + R_8 x^8.
\een 
In new coordinates, two dimensional torus in the $(x^6, x^8)$-direction
will be
of radii $\sqrt{R_6^2 + R_8^2}$ in ${x^6}'$-direction and 
${R_6R_8\over \sqrt{R_6^2 + R_8^2}}$ in ${x^8}'$-direction.
In view of this, we adopt a new notation
\ben
\label{nrad}
R_6' = {\sqrt{R_6^2 + R_8^2}\over 2}, \quad R_8' = {R_6R_8\over \sqrt{R_6^2 + R_8^2}}.
\een
The denominator  $2$ in the definition of $R_6'$ reflects the fact that
the string is stuck  between two D branes  $\pi\sqrt{(R^6)^2
+ (R^8)^2}$ apart. 
Note that  the winding
modes of the open string will be in in the $(x^4, x^5, {x^6}', x^7, {x^8}')$ 
directions and the momenta of the string will be  in the $x^9$ direction
because
D branes impose  Neumann boundary conditions along the $x^9$ direction.

We now evaluate the terms from different sectors. 
We take first the untwisted NS sector. All the traces are taken over the
full Fock space of the open string and includes a sum (integral) over
various various momentum and winding numbers and a sum over Chan Paton 
sectors. If we consider the $D D$ sector, then the summation over the 
winding modes in the $x^{6'}$ directions will involve only integer
winding modes i.e. even-integers in terms of $R_6'$. This is because strings 
start and end on the same D brane and thus cover $R_6'$ an even number of 
times.
If we consider the $D D'$ sector, then we sum over odd-integers in terms of
$R_6'$ because the strings starting and ending on different D-branes 
cover an odd number of times $R_6'$. If we consider both sectors then we 
obtain a summation over all integer winding numbers, thus the term which 
corresponds to winding modes in the $x^{6'}$ direction is
$\theta(R_6')$.  The same discussion holds for the untwisted sector.
If we now consider the twisted R sector, it can come only from strings 
ending on the same brane, so it comes only from the $D D$ sector.
Considering all of the above we obtain:
\ben \label{opennsu}
tr_{NS} (e^{-2t H_o}) &= & A 
(2t)^{-{p\over 2}}
\left({f_3(\tq)\over f_1(\tq)}\right)^8\ta (R_9^{-1}) \prod_{i=4,5,7}
\theta (R_i)\prod_{i=6,8} \theta (R_i')\\
\label{opennst}
tr_{NS} (e^{-2t H_o} (-1)^F\cdot g)& =& - 4A
(2t)^{-{p\over 2}} \left({f_3(\tq)f_4(\tq)\over
f_1(\tq) f_2(\tq)}\right)^4\prod_{i=4,5}\ta (R_i)\\
\label{openrru}
tr_{R} (e^{-2t H_o}) &= &A (2t)^{-{p\over 2}}
\left({f_2(\tq)\over f_1(\tq)}\right)^8 \ta (R_9^{-1})
\prod_{i=4,5,7} \ta (R_i) \prod_{i=6,8} \theta (R_i')\, .
\een 
where $A$ is the normalized $p$-dimensional volume of the brane in
the non-compact directions. 
 In (\ref{opennsu})
and (\ref{openrru}) we use the fact
that we have six compact coordinates and we considered the winding
modes on five of them and the momentum in the  $x^9$ compact direction.
In formula
(\ref{opennst}) we do not have any contribution from the
winding modes on the ${x^6}', x_7, {x_8}', x^9$  directions but we have contribution from
the winding modes on the $x^4, x^5$ directions because they are not acted
upon by $\II$ so they survive the projection. 

We now add the contributions from equations (\ref{c1}) and (\ref{c2})
to obtain the following partition function:
\ben
Z = \int {dt\over 4t} {A}
(2t)^{-{p\over 2}}
\left[\left( {f_4(\tq) \over f_1(\tq)}\right)^8
\ta (R_9^{-1})\prod_{i=4,5,7} \ta (R_i) \prod_{i=6,8} \ta (R_i')
-4 \left({f_3(\tq) f_4(\tq)\over f_1(\tq) f_2(\tq)}\right)^4 
\prod_{i=4,5} \ta (R_i)\right].\nonumber \\
\label{sum2}
\een
Here we used the identity
\ben
f_3(q)^8 - f_2(q)^8 = f_4(q)^8.
\een
We want to know
 when this amplitude becomes zero. First we observe that
the term involving $x^4$ and $x^5$ factors out 
 and therefore there is no conditions for
 the radius of the torus in the  $x^4$ and $x^5$ directions.
For  the other 4 directions,  if
$R_7 = R_6' = R_8 = 1/\sqrt{2}$ and $R_9= \sqrt{2}$
i.e. at the critical radii, then
the contribution of the winding modes and momenta will be 
\be
\label{pdos}
\ta(R_7)\ta(R_9^{-1})
\ta(R_6') \ta(R_8') =  
\left(\sum_{n\in \Zop} \tq^{n^2}\right)^{4} = \left( \sqrt 2 {f_1(\tq) f_3(\tq)\over
f_2(\tq) f_4(\tq)}\right)^{4},
\ee 
where for the last equality we used the sum and the product representation
of the Jacobi $\vartheta$-function $\vartheta_3(0|\tau)$ \cite{embt}
\ben
\vartheta_3(0|\tau) = \sum_{n\in \Zop} q^{n^2}
= \prod_{n=1}^\infty (1 -
q^{2n}) (1 +q^{2n-1})^2
= f_1(q) f_3^2(q),
\een
and the identity
\ben
f_4(q) {1\over \sqrt 2} f_2(q) f_3( q) = 1.
\een
Taking into account our definition (\ref{nrad}), the original critical
radii are $R_6 = R_8 = 1$.

By plugging now (\ref{pdos}) in (\ref{sum2}), we obtain that 
the contribution of the $D D'$ sector to the
partition function is zero at the critical radii $R_7 =
R_6' =  R_8' = 1/\sqrt{2}, R_9 = \sqrt{2}$. 
We see that the critical radius condition has
to be only imposed on the 6,7,8 and 9 directions, the 4 and 5 directions
having arbitrary radii.

We need to check whether the critical radii give a tachyon free model.
We do this by calculating the mass of the NS sector ground state of the
$DD'$ string, this
represents a scalar with $m^2=(R_8^2+R_6^2)/4 -(1/2)$ because the $D$ and 
$D'$ are separated only in these two directions. Thus at the
critical radius $\sqrt{R_8^2+R_6^2}=\sqrt 2$ the mass is zero so there is
no tachyonic scalar. Since there
are no tachyonic fermions, we do not have a tachyonic component.

The critical radii are the ones where we have Bose-Fermi degeneracy at the
massless level. Because besides the compact directions we have 
$(x_1, x_2, x_3)$ non-compact directions, the branes can be at different
positions in those directions. But this just introduces an overall extra
factor of $q^{r^2/2\pi^2}$ in front of the partition function which is
associated with the tension of the open string stretched over a distance
$r$. The potential energy depends on the non-compact direction $r$ and is
equal to the negative of the partition function. It vanish identically for
any value of $r$ and is a monotonically decreasing function of $r$. It is
positive definite because the partition function becomes negative when we
leave the  $R_7 = \frac{1}{\sqrt{2}}, R_6 =  R_8 = 1, R_9 = \sqrt{2}$ 
critical radii case. Therefore for  
 $R_7 > \frac{1}{\sqrt{2}}, R_6 > 1,R_8 > 1, R_9 < \sqrt{2}$, the
interaction between branes is repulsive at all non-compact distances.

We have discussed the case when the $D$ and $D'$ branes are at 
$x^4 = x^5 = 0$. What happens when the D brane is displaced from 
this point to $x^4=b_4, x^5=b_5$ where $b_4$ or $b_5$ is non-zero?
Then the D' must be  displaced to $x^4 =-b_4, x^5 =-b_5$ in order to have
an $\II'$ invariant configuration. 
In this case the mass of the NS sector ground state is 
\be
m^2={ (2 b_4/\pi)^2+(2 b _5/\pi)^2+R_6^2+R_8^2\over 4} -{1\over 2}\, ,
\ee
so the condition of a tachyon free theory is
\be
\label{dep}
4 (b_4/\pi)^2 + 4 (b_5/\pi)^2 + R_6^2 + R_8^2 = 2.  
\ee
Here we assume that $0< b_4 < \pi R_4$ and $0< b_5 < \pi R_5$
But, by considering the winding modes in $(x_4, x_5)$ directions 
from the winding modes in $(x_6, x_7, x_8)$ separately, one can see that
the partition function remains to be the same as in our main case
$(\ref{sum2})$ 
because $\II$ projection
does not act on the $(x_4, x_5)$ directions and again $\II'$ just exchanges
the different sectors. This implies that
$R_6^2 + R_8^2$ must be 
2 if we want the partition function to vanish.
We conclude that $b_4$ $b_5$ must be zero in order to have a tachyon
free theory.

\sectiono{Closed String Approach}

We now compare the results of the previous section with the ones obtained
from  a closed string theory point of view, where the D branes are
viewed as boundary states  \cite{bo1,bo2,bo3,bo4}.
We want firstly to identify the boundary states describing the pair of
D-branes wrapped on one direction of the orbifold. 
Before the $\II'$ projection, the boundary state that represents the
non-BPS brane is of the form~\cite{og2}:
\ben\label{dbound}
|\wt Dp,\ba, \bb,\bc\rangle & = & {1 \over \sqrt{2}}
\left(\mbox{\boldmath $|$}Bp,\ba,\bb,\bc,+
\mbox{\boldmath$\rangle$}_{NSNS}^{U}
    - \mbox{\boldmath $|$}Bp,\ba,\bb, \bc,-
      \mbox{\boldmath$\rangle$}_{NSNS}^{U} \right) \nonumber
\\
& & \quad         + {1 \over 2} \epsilon
\left(\mbox{\boldmath $|$}Bp,\ba,\bb,\bc,+
\mbox{\boldmath$\rangle$}_{RR;(0,\pi R_9)}^{T}
   + \mbox{\boldmath $|$}Bp,\ba,\bb,\bc,-
\mbox{\boldmath$\rangle$}_{RR;(0,\pi R_9)}^{T}\right)\, ,
\een
where we use $(0, \pi R_9)$ to show that we have to make the summation
over the RR components of the twisted sector boundary states located at
$x^9 =0$ and $x^9 = \pi R_9$ respectively. $\epsilon$ is the charge of the
twisted RR sector and can have values $\pm$.
In the above formula
we  represents the coordinates of  the noncompact directions by $\ba$,the $x^4, x^5, x^7$
coordinates by  $\bb$ and the $x^6, x^8$ coordinates by $\bc$ respectively.
We consider only the sector with twisted RR
charge equal to 1. After taking the  $\II'$ projection, we need to
consider a state which is the superposition of two such non-BPS branes 
located at different values of $\bc$, the first one at $\bc_1 = (0, 0)$
and the second one at $\bc_2 = (\pi R_6, \pi R_8)$. 
This state is written as:
\be
\mbox{\boldmath $|$}\wt Dp\mbox{\boldmath$\rangle$} = \mbox{\boldmath $|$}\wt Dp,\ba, \bb = 0,\bc_1 \mbox{\boldmath$\rangle$} + 
\mbox{\boldmath $|$}\wt Dp,\ba, \bb = 0,\bc_2 \mbox{\boldmath$\rangle$}\, ,
\ee
where we consider both branes to be located at $\bb=0$. This state is
invariant under the $\II'$ because the first term is exchanged with the
second one. In order to be able to map the boundary state into itself
under the action of $\II'$ we need to chose the same $\epsilon$ 
parameters for both branes.

We are interested in the tree level amplitude that describes the
exchange of closed strings between two such non-BPS D-branes invariant 
under $\II$ and $\II'$. The amplitude is given by
\be
\label{amp}
\int_{0}^\infty dl \, \langle \wt Dp \mbox{\boldmath $|$}
e^{-l H_c} \mbox{\boldmath $|$} \wt Dp \mbox{\boldmath$\rangle$}
\,,
\ee
where $H_c$ is the closed string Hamiltonian in light cone gauge,
\be
H_c = \pi \vec p^2 + {1\over 4\pi} \vec w^2 + 2\pi
\sum_{\mu=0,3,\ldots,9} \left[
\sum_{n=1}^{\infty} (\alpha^\mu_{-n} \alpha^\mu_{n}
      + \tilde{\alpha}^\mu_{-n} \tilde{\alpha}^\mu_{n})
+ \sum_{r>0} (\psi^\mu_{-r} \psi^\mu_r
      + \tilde\psi^\mu_{-r}\tilde\psi^\mu_{r} ) \right] + 2 \pi C_c\, .
\ee 
The constant $C_c$ is -1 in the untwisted NSNS sector and 0 in the 
twisted RR sector. $\vec p$ and $\vec w$ denote the momentum and winding 
charges as usual.

 In order to calculate the partition function, we need
to introduce the corresponding boundary states in the equation (\ref{dbound}). 

We can now proceed to calculate the tree level closed string amplitude.
Because of the action of $\II'$, the amplitude describing the emission
and re-absorption of closed strings by the D-brane located at 
$\bc_1$ is equal to the contribution of the D-brane located at
$\bc_2$ and we take only
once this contribution. Also, the amplitude describing the emission of a
closed string by the brane at $\bc_1$ and absorption by the brane at
$\bc_2$ is equal with the amplitude describing the inverse process. 
In order to obtain the amplitude, we need to calculate 4 terms, the first
between  $\mbox{\boldmath $|$}\wt Dp,\ba, \bb = 0,\bc_1 \mbox{\boldmath$\rangle$}$ and 
$\mbox{\boldmath $|$}\wt Dp,\ba, \bb = 0,\bc_2 \mbox{\boldmath$\rangle$}$, the second one between
$\mbox{\boldmath $|$}\wt Dp,\ba, \bb = 0,\bc_2 \mbox{\boldmath$\rangle$}$ and $\mbox{\boldmath $|$}\wt Dp,\ba, \bb = 0,\bc_1
\mbox{\boldmath$\rangle$}$, the third one between 
$\mbox{\boldmath $|$}\wt Dp,\ba, \bb = 0,\bc_1 \mbox{\boldmath$\rangle$}$ and itself and the fourth between
$\mbox{\boldmath $|$}\wt Dp,\ba, \bb = 0,\bc_2 \mbox{\boldmath$\rangle$}$
and itself. We add all these terms and then divide by 2 because of the
$\II'$ projection.

We need to identify the untwisted and twisted sectors.
As any orbifold discussion, the twisted sectors appears at fixed points of 
the orbifold. Because the D-branes are at fixed points of $\II$, the
twisted sectors are obtained from the amplitude of emission and absorption
by
the same brane and we only take the contribution once, as discussed
before. The untwisted sector is obtained from both the amplitude of
emission by
one brane and absorption by the other brane and from the amplitude of
emission and absorption by the same brane.
Another difference which arises in our case is that we need to have
in the twisted sector the winding modes coming from the two compact
directions $x^4$ and $x^5$ that $\II'$ is not acting upon. Then
the contribution from the twisted RR sector will be
\be
\frac{1}{2}  \widetilde{\cal N}^2  \left({f_2(q) f_3(q) \over f_1(q) f_4(q)}\right)^4
\left(\prod_{i=4,5}
\sum_{m\in\Zop} e^{-l\pi m^2/R_i^2}\right), \quad q =e^{-2\pi l}\, ,
\label{twRR}
\ee 
which is obtained from 
\ben
\frac{1}{4}\mbox{\boldmath $\langle$} Bp,\ba,\bb,\bc_1,\pm
\mbox{\boldmath $|$}_{RR}^{T}\quad e^{-l H_c}
\mbox{\boldmath$|$}Bp,\ba,\bb,\bc_1,\pm\mbox{\boldmath$\rangle$}_{RR}^{T} 
\nonumber \\
+\frac{1}{4} \mbox{\boldmath $\langle$}
Bp,\ba,\bb,\bc_2,\pm\mbox{\boldmath
$|$}_{RR}^{T}\quad e^{-lH_c} \mbox{\boldmath
$|$}Bp,\ba,\bb,\bc_2,\pm \mbox{\boldmath$\rangle$}_{RR}^{T}.
\een
Here the factor $1/4$ comes from the fact that the charge of the twisted RR sector
 $\epsilon$  in the boundary state
decomposition (\ref{dbound})  are equal for the two
branes. $\widetilde{\cal N}^2 $ in (\ref{twRR}) is a normalization constant
to be determined later.

In order to calculate the untwisted sector contribution to the amplitude,
we need to remember the form for the states appearing in the untwisted RR
sector. They are:
\be
\label{boundaryNSNS}
|Bp,\ba,\bb,\bc,\eta\rangle_{NSNS}^U = {\cal N} \int
\left(\prod_{\mu {\mbox{ transverse}}}
 dk^\mu e^{i \bk \cdot \ba} \right)
\left(\prod_{i=6,8} \sum_{m\in\Zop}
e^{i m c_i/R_i} \right)  
\wh{|Bp,\bk,\bm,\eta\rangle}_{NSNS}^U, 
\ee
where $\bk$ denotes the momentum in the non-compact directions, and
$m/R_i$ the momentum along the $i$th compact direction and
${\cal N}$ is a normalization factor to be determined later. 
Here we are taking integration  over the directions transverse to
the D-brane and 
$\wh{|Bp,\bk,\bm,\eta\rangle}_{NSNS}^U$ denotes the coherent momentum
eigenstate 
\ben\label{bound}
\wh{|Bp,\bk,\bm,\eta\rangle} 
& = & \exp \left(
\sum_{n=1}^\infty \left[
- {1\over n} \sum_{\mu\in\C} \alpha^\mu_{-n} \tilde\alpha^\mu_{-n}
+ {1\over n} \sum_{\mu\in\hat\D} \alpha^\mu_{-n} \tilde\alpha^\mu_{-n}
\right] \right. \nonumber \\
& & \qquad\qquad\left. + i \eta \sum_{r>0} \left[
- \sum_{\mu\in\C} \psi^\mu_{-r} \tilde\psi^\mu_{-r}
+ \sum_{\mu\in\hat\D} \psi^\mu_{-r}
\tilde\psi^\mu_{-r} 
\right] \right) \wh{|Bp,\bk,\bm,\eta\rangle}^{(0)},\,\,\,\,\,
\een   
where  $\wh{|Bp,\bk,\bm,\eta\rangle}^{(0)}$ 
denotes the Fock vacuum labeled by
the quantum numbers $\bk,\bm$.                                                                    
When we plug the
states into equation (\ref{amp}), we will have a summation of the following term:
\be
1 + e^{i \pi m} + e^{-i \pi m} + 1 = 4 \cos^{2} (\pi m/2).
\ee
The terms in the left hand sides come from the closed strings emitted and
absorbed by the same D-brane (first and fourth) and from the closed
strings emitted and absorbed by different branes situated at 0 and $2 R_6'$
where $R_6'$ is defined in (\ref{nrad}). This is non-zero only for
even-integer values for $m$, therefore the summation over the winding modes
on the $x_6'$ direction will be over even-integer numbers.
Then the contribution of the untwisted sector to the amplitude is (after
taking the $\II'$ projection) is: 
\ben\label{amp1}
\prod_{i=4,5,7} \sum_{m\in\Zop}
e^{-l\pi m^2/R_i^2} ( \sum_{m \in\Zop}
e^{-l \pi m^2 R_9^2})
(\sum_{m\in 2 \Zop} e^{-l\pi m^2/4 R_6'^2}) (\sum_{m\in \Zop}
e^{-l\pi
m^2/R_8'^2}) {f_3^8(q) - f_4^8(q) \over f_1^8(q)} 
\een
where  $q =e^{-2\pi l}$. 

By adding up all the contributions to the amplitude, we obtain
\ben\label{amp2bis}
{1\over 2} \int_0^\infty dl\, l^{-{4-p \over 2}} 
\left[ 2 {\cal N}^2 \prod_{i=4,5,7} \sum_{m\in\Zop}
e^{-l\pi m^2/R_i^2}\sum_{m \in\Zop}
e^{-l \pi m^2 R_9^2}\sum_{m\in 2 \Zop} e^{-l\pi m^2/4 R_6'^2}
\times\right. \nonumber \\
\times \left. \sum_{m\in \Zop} e^{-l\pi m^2/R_8'^2}  {f_3^8(q) -
f_4^8(q) \over f_1^8(q)}
- \widetilde{\cal N}^2 \left( {f_2(q) f_3(q) \over f_1(q) f_4(q)}
\right)^4\prod_{i=4,5} \sum_{m\in\Zop} e^{-l\pi m^2/R_i^2}\right], \
\een
where  $q =e^{-2\pi l}$ and
 $\cal N, \widetilde{\cal N}$ are the normalization constants
introduced for the boundary states. The power of $l$ comes from the fact
that the D-brane is extended in $p$ of the 4 non-compact directions. 

To determine the normalization constants, we apply a modular 
transformation $t = 1/2 l$ which changes
 the closed string tree  amplitude
into an open string 1-loop amplitude.
Using the properties of the $f_i$ functions,
\be\label{trans}
\begin{array}{rclrcl}
f_1(e^{-\pi/t}) & = & \sqrt{t} f_1(e^{-\pi t})\,, \qquad &
f_2(e^{-\pi/t}) & = & f_4(e^{-\pi t})\,, \\
f_3(e^{-\pi/t}) & = & f_3(e^{-\pi t})\,, \qquad &
f_4(e^{-\pi/t}) & = & f_2(e^{-\pi t})\,,
\end{array}
\ee
together with the identity
\be \label{identity}
\sum_{m\in\Zop} e^{-\pi l (m/R)^2} = {R \over \sqrt{l}}
\sum_{m\in\Zop} e^{-2t\pi (mR)^2} = R\sqrt{2t} \sum_{m\in\Zop}
q^{2m^2 R^2}, \quad q = e^{-\pi t}\, ,
\ee
we can express (\ref{amp2bis}) as 
\ben
\label{closamp1}
{1\over 2} \int_0^\infty {dt\over 2t} t^{-{p\over 2}}
 2^{6-p\over 2}R_4 R_5  
\prod_{i=4,5} \theta(R_i)      
\left[ 8 {\cal N}^2 \frac{R_7 R_6' R_8'}{R_9} 
 \theta(R_7)    \theta(R_9^{-1}) \prod_{i=6,8} \theta(R_i') 
{f_3^8(q) - f_2^8(q) \over f_1^8(q)}\right.   \nonumber \\  
\left.- \widetilde{\cal N}^2 \left({f_3(q)f_4(q) \over
f_1(q) f_2(q)}\right)^4\right],\,\,
\een
where $q = e^{-\pi t}$.
By  comparing the form of the above  partition function 
 with the open string theory partition function
(\ref{sum2}),  we set  the values of normalization constants 
${\cal N}^2$ and $\widetilde{\cal N}^2$ to be 
\ben
64 \frac{R_4R_5R_7 R_6' R_8'}{R_9}{\cal N}^2 = A\, ,\\
8 R_4R_5\widetilde{\cal N}^2 = 4A\, ,
\een
so that the  closed string amplitude becomes 
\ben\label{closamp2}
{A\over 4} \int_0^\infty dl l^{-{4-p\over 2}}{1\over R_4 R_5}
(\prod_{i=4,5} \sum_{m\in\Zop} e^{-l\pi m^2/R_{i}^2})  
\left[\frac{R_9}{16 R_7 R_6' R_8'}
(\sum_{m\in\Zop} e^{-l\pi m^2/R_7^2})
(\sum_{m\in\Zop} e^{-l\pi m^2 R_9^2})  \times \right. \nonumber \\ 
\left.  (\sum_{m\in 2 \Zop} e^{-l\pi m^2/4 R_6'^2})
(\sum_{m\in \Zop} e^{-l\pi m^2/R_8'^2)})
{f_3^8(q) - f_4^8(q) \over f_1^8(q)}
- \left({f_2(q) f_3(q) \over  f_1(q) f_4(q)}\right)^4\right]. \
\een
If we choose $R_6'  = R_8' = R_7 =
1/\sqrt{2}$,
$R_9 = \sqrt{2}$, then the partition function becomes zero. The
corresponding values for $R_6, R_8$ are $R_6 = R_8 = 1$.
So, by using the boundary state formalism we found  critical
radii for which  the tree level partition function vanishes.

Things might change if one takes higher-loop corrections in both open
string and closed string theories. 

\vskip .5cm

{\bf Acknowledgments}

We would like to thank M. Gaberdiel and especially A. Sen for insightful
 suggestions and  comments. The work of K. Oh is supported in part by
NSF grant PHY-9970664. K. Oh thanks Department of Mathematics at
University of California, Santa Barbara for its hospitality
during his stay.

\end{document}